\documentclass{svjour3}                     

\smartqed 
\usepackage[utf8]{inputenc}
\usepackage{amsmath}
\usepackage{amsfonts}
\usepackage{amssymb}
\usepackage{listings}
\usepackage{mathrsfs}
\usepackage{times}
\usepackage{float}
\usepackage{color}
\usepackage{setspace}
\usepackage{color,soul}
\usepackage{caption}

\usepackage{amsmath,amsfonts,amsthm,eucal, bm}
\usepackage{enumerate}
\usepackage[letterpaper,margin=3cm]{geometry}
\usepackage{float}
\usepackage[title]{appendix}
\usepackage{color}
\usepackage{comment}
\usepackage{multirow}
\usepackage{siunitx}
\usepackage{pgfplots}
\pgfplotsset{compat=1.3}

\usepackage{subcaption}
\usepackage{graphicx}
\usepackage{longtable} 
\usepackage{graphicx}  
\usepackage{calc}      

\usepackage[
pdfstartview=XYZ,
bookmarks=true,
colorlinks=true,
linkcolor=blue,
urlcolor=blue,
citecolor=blue,
pdftex,
bookmarks=true,
linktocpage=true,   
hyperindex=true
]{hyperref}

\usepackage{lipsum}
\usepackage{lineno}

\usepackage[numbers, sort&compress]{natbib}

\usepackage{graphicx}
\usepackage{pstool}
\usepackage{psfrag}
\usepackage{epstopdf}
\usepackage{booktabs} 
\usepackage{lineno}
\usepackage{import}
\usepackage[export]{adjustbox}

\graphicspath{{figures/}}
\usepackage{fancybox}

\def\newbullet{%
  \unitlength 1ex%
  \begin{picture}(1,1)
  \put(0.5,0.6){\circle*{0.4}}
  \end{picture}}

\newcommand{\Four}[1]%
           {\hbox{$\stackrel{4}{{}_{\vphantom{S}}{#1}_{\vphantom{S}}}$}{}}

\usepackage{algorithm}
\usepackage[noend]{algpseudocode}

\journalname{My-journal}
\begin{document}

\title{Embedding an ANN-Based Crystal Plasticity Model into the Finite Element Framework using an ABAQUS User-Material Subroutine}
\titlerunning{Embedding an ANN-Based Crystal Plasticity Model in ABAQUS UMAT}        


\author{Yuqing He$^{1}$       \and
        Yousef Heider$^{1,2,\ast}$        \and
        Bernd Markert$^{1}$      
}


\institute{$^1$Institute of General Mechanics, 
               RWTH Aachen University. \\
               Eilfschornsteinstr. 18, 52062 Aachen, Germany\\[1mm]
$^2$ Institute of Mechanics and Computational Mechanics (IBNM), Leibniz University Hannover.\\
Appelstr. 9A, 30167 Hannover, Germany\\[1mm]
$^{\ast}$ Corresponding author: Yousef Heider.\\ 
\email{yousef.heider@ibnm.uni-hannover.de}
}

\date{\today}

\maketitle

\begin{abstract} 

This manuscript presents a practical method for incorporating trained Neural Networks (NNs) into the Finite Element (FE) framework using a user material (UMAT) subroutine. The work exemplifies crystal plasticity, a complex inelastic non-linear path-dependent material response, with a wide range of applications in ABAQUS UMAT. However, this approach can be extended to other material behaviors and FE tools. The 
use of a UMAT subroutine serves two main purposes: (1) it predicts and updates the stress or other mechanical properties of interest directly from the strain history; (2) it computes the Jacobian matrix either through backpropagation or numerical differentiation, which plays an essential role in the solution convergence. By implementing NNs in a UMAT subroutine, a trained machine learning model can be employed as a data-driven constitutive law within the FEM framework, preserving multiscale information that conventional constitutive laws often neglect or average. 
The versatility of this method makes it a powerful tool for integrating machine learning into mechanical simulation. While this approach is expected to provide higher accuracy in reproducing realistic material behavior, the reliability of the solution process and the convergence conditions must be paid special attention. While the theory of the model is explained in \cite{HEIDER2020}, exemplary source code is also made available for interested readers [\href{https://doi.org/10.25835/6n5uu50y}{https://doi.org/10.25835/6n5uu50y}].

\keywords{Machine learning \and  Path-dependent material behavior \and Crystal plasticity \and Finite element \and User material subroutine \and UMAT \and ABAQUS}

\end{abstract}

\begin{center}
  \fbox{ 
    \begin{minipage}{0.9\textwidth}
    \textbf{Abbreviations} 
    \begin{longtable}{ll ll}
ANN & Artificial Neural Network & UMAT & User Material Subroutine \\ 
FE  & Finite Element            & FEM  & Finite Element Method \\ 
CNN & Convolutional Neural Network & DSA & Dynamic Strain Aging \\ 
LSTM & Long Short-Term Memory & RNN & Recurrent Neural Network \\ 
DDSDDE & Material Jacobian Matrix & HDF5 & Hierarchical Data Format Version 5 \\ 
ML & Machine Learning & IBVP & Initial Boundary Value Problem \\ 
f.c.c. & Face-Centered Cubic & FFNN & Feed-Forward Neural Network \\
    \end{longtable}
    \end{minipage}
  }
\end{center}

\section{Introduction}
\label{sec:intro}
In solid mechanics, the conventional constitutive laws explicitly formulate the stress-strain relations with symbolic expressions. Such relations are mostly constructed from simplified assumptions and based on empirical observations. Thus, the microscopic information is partially or entirely neglected from the material models \cite{GIOVINE2022}. However, data-driven constitutive models make it possible to describe the stress-strain relations merely based on acquired data, without specifying a certain constitutive model, see, e.g., \cite{HEIDER2020, Heider2021_Habil, HeiderSun2023_PAMM_PlasObj}. Machine learning (ML) techniques can be applied to capture such constitutive relations. Hence, integrating ML models as a substitution of model-based constitutive laws into the FE framework offers the potential to incorporate lower-scale information, without explicitly specifying any material parameter in a model. In this regard, \citet{LEFIK2003} used ML to model elasto-plasticity and biaxial non-linear behavior, whereas \citet{OISHI2017} used ML for the stiffness matrix in FEM. \citet{TAO2021} utilized ML to fit the engineering constants of the constituents of a fiber-reinforced composite, as well as the progressive damage constitutive law. \citet{XU2021} built the Cholesky-factored symmetric positive definite neural network to predict constitutive relations in dynamic equations. \citet{PANTIDIS2023} employed a ML model that utilized material point deformation as input and produced element Jacobian matrix and residual vector as output. In other works, like by \citet{HASHASH2004,HUANG2020a,LI2019},  ML models are trained to acquire the material's constitutive laws. These models utilized strain components as input and stress components as output.
In the work of \citet{KoeppeBamerMarkert2020_An_Intelligent}, a hybrid substructuring approach is introduced using the TRUNet deep neural network to develop a nonlinear, inelastic intelligent meta element that accelerates finite element simulations in history-dependent mechanical problems with reduced computational effort.
While the focus of the current work is on regression models, convolutional neural networks (CNN), as a class of ANN, are also widely used within multiscale material modeling to generate ML-based constitutive models.  CNNs require mostly data that have a grid-like nature, like images or time-series data, see, e.g., \cite{GU2018CNN, EIDEL2023_CNN, Aldakheel_2023_CNN_Magnetostatics,Dhillon_Verma_2020_reviewCNN,Aldakheel_2023_CNN_Hetero, Tandale_2023_CNN_RNN, ChaabanEtAl2023_ML_LBM, HeiderAldakheelEhlers_2024CNN} for review and applications. 


Several research groups worked on the implementation of ML material models within the Finite Element Method (FEM) as an alternative to traditional constitutive models. To mention some, \citet{Guan2023_FEM_ML} implemented their  ML constitutive model within multi-scale granular materials modeling. In this, the FEM and the Discrete Element Method (DEM) are coupled to generate training samples. They showed that the FEM-ML framework offers considerable improvements in terms of computational efficiency and the ability to simulate the mechanical responses of granular materials. In the work of \citet{Suh2023_PyTorch_ABAQUS_Umat}, an open-access PyTorch-ABAQUS deep-learning framework for a group of plasticity models is presented. In this, an interface code is presented in which the weights and biases of the trained NNs are automatically converted into a FORTRAN code that is compatible with the UMAT/VUMAT of ABAQUS. In connection with unsaturated porous media, \citet{HeiderHSSuh2020_ML_offline} implemented a trained ML material model of retention curve within a Python FE framework. They discussed through numerical examples the accuracy improvement due to ML material models as well as the instability sources.
In the work of \citet{WeberWagnerFreitag2023}, challenges associated with the application of ANN elasto-plasticity material models within FE calculations together with implementation instructions are presented. As a remedy for some challenges, they proposed an approach that includes constrained neural network training.
An application of ANN in modeling Dynamic Strain Aging (DSA) in stainless steel and the implementation of the developed ANN-based model in the FE formulation is presented by \citet{PatraEtAl2023_FE_ANN}.
In the work of \citet{LourencoEtAl2022_VMAT_met12030427}, the use of ML techniques for improving the accuracy of material constitutive models in metal plasticity using ABAQUS VUMAT is discussed. 
In \citet{JANG2021102919_UMAT}, a J2-plasticity constitutive model to predict elastoplastic behavior is proposed, where the ML model is implemented in ABAQUS UMAT. The ML model is used only for nonlinear plastic loading while leaving linear elastic loading and unloading to the physics-based model. They reported that the results of their model are in good agreement with those from the conventional constitutive models for a given boundary-value problem.
For multiscale modeling, \citet{TeijeiroEtAl2021_ML_FEM} proposed a methodology that combines FEM with ANN in the numerical modeling of systems with behavior that involves a wide span of spatial scales.
In the context of biomechanics, \citet{Lee2021_ML_FEM} discussed the potential of using the ML-based approach in solving FE-based classification questions.

While feed-forward neural networks (FFNN) can be used to train path-independent material models (e.g., elasticity), recurrent neural
networks (RNN) are used to capture responses that depend on the deformation history, i.e. path- or time-dependent responses. This includes for instance,  viscoelasticity and elasto-plasticity, see, e.g., \cite{LAZAR2009,GORJI2020,MAIA2023,YUAN2018,MIYAZAWA2019,HEIDER2020}. In this context, the Long Short-Term Memory (LSTM) neural network, an integral component of the RNN paradigm introduced by \citet{HOCHREITER1997}, is considered in this work. 
%
In this manuscript, we build upon the model presented by \citet{HEIDER2020}, which introduced a graph-based neural network model focusing on the material point response in crystal plasticity. This model is extended later in \cite{HeiderSun2023_PAMM_PlasObj} to fulfill in addition to objectivity also the dissipation inequality.
Our current work specifically proposes a methodology that seamlessly integrates LSTM neural networks into the FEM framework of ABAQUS. The goal is to employ the trained machine learning constitutive model as an alternative to the traditional crystal plasticity material model by using the user-defined material (UMAT) subroutine of ABAQUS.
In summary, the highlights of this manuscript can be summarized in the following points:  
\begin{itemize}
    \itemsep0.5em
    \item[\newbullet] The primary focus is on the history-dependent or path-dependent material mechanical behaviors, such as those encountered in crystal plasticity.
    \item[\newbullet] An approach is proposed to facilitate the integration of machine-learning-based material models into FE analysis. In this work, the simulations are conducted in ABAQUS.
    \item[\newbullet] This paper adopts the UMAT subroutine designed especially for single-grain crystal plasticity materials as the test platform.
    \item[\newbullet] This subroutine inherently ensures the fulfillment of equilibrium equations and employs the Newton-Raphson method for convergence, augmenting the accuracy and efficiency of the simulation outcomes.
\end{itemize}

\section{Theoretical fundamentals}
\label{sec:Theory}

\subsection{Data-based ML plasticity model and data preparation}
\label{sec:PlasticityModel}

\subsubsection{Overview and procedures}
\begin{figure}[h!]
    \centering
    \includegraphics[width=0.95\textwidth]{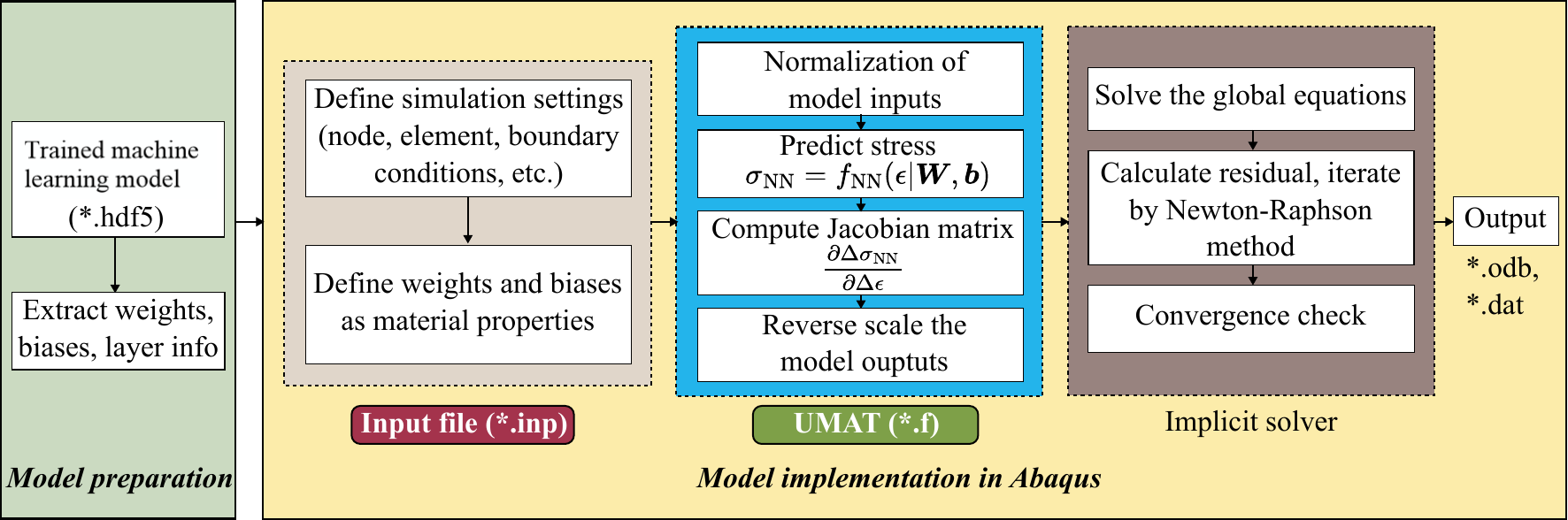}
    \caption{Illustration of the process flow: From model preparation to implementation in ABAQUS UMAT. The input file and the UMAT subroutine file can be modified within the ABAQUS components, while the implicit solver is the standard integral ABAQUS component.}
    \label{fig:ABAQUS_flow}
\end{figure}
As illustrated in Fig. \ref{fig:ABAQUS_flow}, integrating a data-driven material model into a FE solver starts with building, training, and validating the neural networks (NNs) using pre-generated datasets. Details and references regarding dataset generation and training settings are presented in Section~\ref{sec:generation_dataset}. 
Subsequently, the trained models are saved in Hierarchical Data Format version 5 (HDF5) files, which encompass the model architecture (weights, biases) and hyperparameters (number of hidden layers, learning rate, activation function, etc.). These parameters can be transferred to the ABAQUS interface via user-defined material constants. Notably, this process is implemented in the Python programming language with the libraries of Tensorflow \cite{abadi2016tensorflow} and Keras version 2.12.0 \cite{chollet2015keras}. 

To configure the simulation, an ABAQUS input file is used to specify or overwrite various parameters within the simulation environment. These parameters include mesh nodes and elements, loads, boundary conditions, as well as initial conditions. 
Within ABAQUS, the UMAT subroutine customizes the material responses by performing two functions: 
\begin{enumerate}
    \item  It predicts and updates the stresses \textbf{STRESS(NTENS)} on a material point based on the normalized strain history using the forward path of the trained LSTM neural network. For general three-dimensional (3D) problems, the symmetric stress tensor comprises NTENS=6 components.
    \item It computes the Jacobian square matrix \textbf{DDSDDE\,(NTENS, NTENS)} representing the 2D format of the material Jacobian matrix with NTENS rows and NTENS columns. Each element $\displaystyle\text{DDSDDE\,(I, J)}=\frac{\partial \Delta \sigma_{I}}{\partial \Delta \epsilon_{J}}$ signifies the change in the $I^\text{th}$ stress component caused by an infinitesimal perturbation of the $J^\text{th}$ component of the strain increment array. DDSDDE can be computed either through NN's backward propagation or numerical differentiation. Due to the path dependency of the LSTM neural network, DDSDDE is continually updated using previous stresses and strains. 
\end{enumerate}
The material constants \textbf{PROPS(NPROPS)}, which represent the previously extracted NN's settings, are defined in the ABAQUS input file. These constants enable ABAQUS to incorporate the trained neural network into the solving process during simulation.

Since the ABAQUS implicit solver uses unscaled values for iteration, both predicted variables must be reverse-scaled into their original value ranges. Following the approximation of a converged solution with the ABAQUS implicit solver, the simulation results are exported as *.odb or *.dat files for subsequent postprocessing and plotting.

\subsubsection{Generation of the training datasets}
\label{sec:generation_dataset}
The synthetic datasets are generated by subjecting a face-centered cubic (f.c.c.) crystal to stepwise increasing strain-controlled loadings, resulting in strain-stress material responses. 
The overall crystal stress is generated by rotating the slip systems based on Borja's ``ultimate algorithm" \cite{BORJA2013}, which is a stable solution of the micro-scale crystal plasticity model. 
In particular, an f.c.c crystal with a maximum of 12 activatable slip systems at 49 different configurations is subjected to strain-controlled loading, yielding a history of strain-stress components. A total of 1176 experiments were conducted, comprising 24 numerical experiments for each of the 49 crystal configurations. The crystal was subjected to monotonic incremental loading of one strain component during each numerical experiment. Specifically, 100 increments were applied for each experiment, which allowed capturing the history-dependent constitutive relation up to 100 history steps. The dataset generation process should ensure an even distribution of data points throughout the entire range of interest. The material parameters and the loading cases are listed in the preceding research of the second author \cite{HEIDER2020}. For training purposes, the database is also divided into training, validation, and testing subsets. The generated database input and output are also scaled into the range [0,1] using the MinMaxScaler object from the scikit-learn.preprocessing library. 
%

\begin{figure}[h!]
    \centering
    \includegraphics[width=0.5\textwidth]{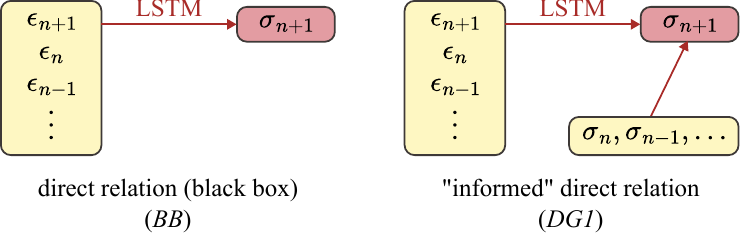}
    \caption{Directed and informed graph representing the information flow in the machine learning-based crystal plasticity material model based on \cite{HeiderHSSuh2020_ML_offline}.}
    \label{fig:bb_dg1}
\end{figure}

As illustrated in Fig.~\ref{fig:bb_dg1}, two distinct graphs are introduced in \cite{HeiderHSSuh2020_ML_offline} to predict the current stress from the strain history with different input structures. 
%
The ``black box'' graph relies only on the strain history, while the ``informed" approach incorporates additional intermediate inputs, such as the previously predicted stresses. The input dimension of the neural network for the ``black box" approach is (6, $l_\text{sequence}$), while for the ``informed" approach it is (12, $l_\text{sequence}$). \\
When implementing the UMAT subroutine, the solution-dependent state variables \textbf{STATV(NSTATV)} facilitate the temporary retention of stress and strain history. This enables the historical information to be available for the current calculation step, thereby reproducing the input of the neural networks. Within the current step $n+1$, ABAQUS passes the strain components from step $n-2$ to step $n-2-l_{\text{sequence}}$ as the state variable, while the strain from the last two steps can be directly computed from the previous strain $\epsilon_{n}$ and the strain increment $\Delta \epsilon_{n+1}$, where $l_{\text{sequence}}$ is the input's length of interest.

\subsection{Mathematical expression for forward and backward pass in a single LSTM memory cell}
\label{sec:math_lstm_cell}
\begin{figure}[h]
    \centering
    \includegraphics[width=0.5\textwidth]{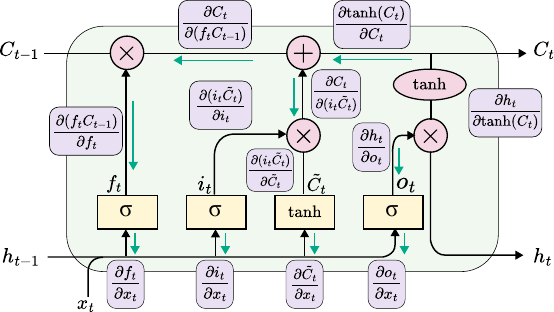}
    \caption{Diagram illustrating the backward pass of the information flow within a single LSTM memory cell, with labeled derivatives on each edge.}
    \label{fig:backward_lstm_cell}
\end{figure}

\citet{HOCHREITER1997}  first introduced long-short-term memory (LSTM) in \num{1997} as a variation of the recurrent neural networks (RNNs). The primary objective of LSTM is to mitigate the vanishing gradient problem commonly encountered in RNNs. This is achieved through the utilization of gates that dynamically control the memory. To elucidate the functioning of the LSTM layer, a visual representation of a single LSTM cell is provided in Fig.~\ref{fig:backward_lstm_cell}. An LSTM layer comprises multiple such cells. Each cell receives three primary inputs: an external current input $x_t$ from the sequence, the internal hidden state $h_{t-1}$, and the cell state $C_{t-1}$ from the preceding time step. Subsequently, the cell updates the hidden state $h_t$ and cell state $C_t$. The final output of the memory cell is encapsulated by the hidden state computed at the last step, which is denoted as $h_{N}$ for $N$ steps in total.
Additionally, within the LSTM cell, there exist three fundamental gates. The input gate ($i_t$) regulates the inflow of information into the memory, the forget gate ($f_t$) controls which information is retained in the memory, and the output gate ($o_t$) determines the extent to which the output is influenced by each state respectively. 

This work specifically employed the LSTM layer from the Tensorflow Keras package to construct neural networks. The mathematical expressions presented in this context align with Keras' implementation and should remain consistent if any other deep learning framework is chosen for the same purpose. 
For each LSTM cell within the framework of Keras, we use two distinct sets of weight matrices: (1) the kernel weight matrix $\boldsymbol{W}_{\text{kernel}}$, which is responsible for weighting the input sequence, and (2) the recurrent weight matrix $\boldsymbol{W}_{\text{recurrent}}$, which is responsible for weighting the preceding time step. These matrices consist of four vertical submatrices corresponding to the input gate, forget gate, output gate, cell and cell input activation vectors, i.e. 
$$
\boldsymbol{W}_{\text{kernel/recurrent}}=\left\{ \boldsymbol{W}_{\text{kernel/recurrent}}^{i} | \boldsymbol{W}_{\text{kernel/recurrent}}^{f} | \boldsymbol{W}_{\text{kernel/recurrent}}^{C} | \boldsymbol{W}_{\text{kernel/recurrent}}^{o} \right\}
$$ with one set of biases, i.e. $\boldsymbol{b} = \left\{ 
\boldsymbol{b}_i | \boldsymbol{b}_f | \boldsymbol{b}_C | \boldsymbol{b}_o \right\}$.

As a starting point of calculation, the forget gate $f_{i}^{(t)}$ for time step $t$ and cell $i$ is computed as introduced in \cite{HOCHREITER1997}:
\begin{equation}
f_i^{(t)} = \sigma \left( \sum_{j} W_{\text{kernel},i,j}^{f}\, x^{(t)}_j + \sum_{j} W_{\text{recurrent},i,j}^{f}\, h_{j}^{(t-1)} + b_i^{f} \right) \,.
\end{equation}
We also have the following expression to update the cell state: 
\begin{align}
\Tilde{C_i}^{(t)} = & \text{tanh}\left( \sum_j W_{\text{kernel},i,j}^{C}\, x_{j}^{(t)} + \sum_j W_{\text{recurrent},i,j}^{C}\, h_{j}^{(t-1)} + b_i^C \right), \\[1ex]
C_i^{(t)} =& f_i^{(t)} * C_i^{(t-1)} + i^{(t)}_i * \Tilde{C_i}^{(t)}.
\end{align}
The calculation of the output gate is
\begin{equation}
o_i^{(t)} = \sigma \left( \sum_j W_{\text{kernel},i,j}^{o} \, x_j^{(t)} + \sum_j W_{\text{recurrent},i,j}^{o}\, h^{(t-1)}_j + b_i^o \right).
\end{equation}
The update of the hidden state is
\begin{equation}
h_i^{(t)} = o_i^{(t)} * \text{tanh}\left(C_i^{(t)}\right).
\end{equation}


\subsection{Numerical differentiation}
Numerical differentiation is a computational method used for calculating derivatives. Unlike the backpropagation approach, numerical differentiation does not depend on the network's architecture, making it more applicable to complex network structures. The underlying concept of numerical differentiation is straightforward, wherein the original derivative is substituted with the derivative of an interpolation polynomial, which is generated with the help of a Taylor expansion \cite{DAHMEN2008}. In this work, the central difference method is used to determine the optimal discretization distance. 

Considering the vectorial representation of the stress tensor $\boldsymbol{\sigma}$ and the strain tensor $\boldsymbol{\epsilon}$, the central difference method delivers the partial derivative of the stress component $\sigma_{i}(\boldsymbol{\epsilon})$ with respect to the strain component $\epsilon_{j}$ as 
\begin{equation}
    \displaystyle
    \frac{\partial \sigma_{i}\left(\boldsymbol{\epsilon}\right)}{\partial \epsilon_{j}} = \underbrace{\frac{\sigma_{i} \left(\boldsymbol{\epsilon} + \frac{1}{2}\Delta \epsilon_{j}\right) - \sigma_{i}\left(\boldsymbol{\epsilon}-\frac{1}{2}\epsilon_{j}\right)}{\Delta \epsilon_{j}}}_{\text{Explicit formula}} - \underbrace{\frac{\Delta \epsilon_j^2}{24}\frac{\partial \sigma_{i}^{3}}{\partial^{3} \epsilon_j}}_{\text{Remainder term}}.
\end{equation}
The error introduced by numerical differentiation comes from the truncation and rounding errors. Let the absolute error, which indicates the resolution of approximation, be \num{e-9}, the optimal discretization size would be then $\Delta \epsilon_{j} \approx 10^{-2}$. 
Notice that the numerical differentiation method employed in this study involves using the scaled input and scaled output data to calculate the derivative (see Section \ref{sec:generation_dataset}). Consequently, it is necessary to reverse-scale the Jacobian matrix back to its original value range: 
\begin{align}
\displaystyle
\frac{\partial \sigma_{i, \text{scaled}}\left(\boldsymbol{\epsilon}\right)}{\partial \epsilon_{j, \text{scaled}}} =\,&  \frac{\partial \left(\displaystyle \frac{\sigma_{i} - \sigma_{i, \text{min}}}{\sigma_{i, \text{max}} - \sigma_{i, \text{min}}} \right)}{\partial \left(\displaystyle \frac{\epsilon_{j}-\epsilon_{j,\text{min}}}{\epsilon_{j,\text{max}}-\epsilon_{j, \text{min}}} \right)} 
=\, \frac{\epsilon_{j,\text{max}} - \epsilon_{j,\text{min}}}{\sigma_{i,\text{max}} - \sigma_{i,\text{min}}} \frac{\partial \left( \sigma_{i}\left(\boldsymbol{\epsilon}\right) - \sigma_{i, \text{min}} \right)}{\partial \left( \epsilon_{j} - \epsilon_{j, \text{min}} \right)} 
=\, \frac{\epsilon_{j,\text{max}} - \epsilon_{j,\text{min}}}{\sigma_{i,\text{max}} - \sigma_{i,\text{min}}} \frac{\partial \sigma_{i}\left(\boldsymbol{\epsilon}\right)}{\partial \epsilon_{j}} \notag\\[1ex]
\quad\Rightarrow \quad\frac{\partial \sigma_{i}\left(\boldsymbol{\epsilon}\right)}{\partial \epsilon_{j}} = & \left( \frac{\sigma_{i,\text{max}} - \sigma_{i,\text{min}}}{\epsilon_{j,\text{max}} - \epsilon_{j,\text{min}}} \right) \frac{\partial \sigma_{i,\text{scaled}}\left(\boldsymbol{\epsilon}\right)}{\partial \epsilon_{j,scaled}}, 
\end{align} \\
where $\sigma_{i, \text{max}/\text{min}}$ and $\epsilon_{j, \text{max}/\text{min}}$ are the maximal and minimal constant values each of the stress and strain components. Using the range of data values from the generated database, the error resulting from numerical differentiation is estimated to be about \num{3.16e-05}. Since the Jacobian matrix cannot be defined exactly in a numerical context, to obtain an acceptable converged solution, the largest residual in the balance equation $r_{max}^{\alpha}$ from the tolerance definition should not be greater than the error of the numerical differentiation. 

\section{Numerical examples} 
\label{sec:NumExamples}
\begin{figure}[!h]
    \centering
        \includegraphics[width=0.45\textwidth]{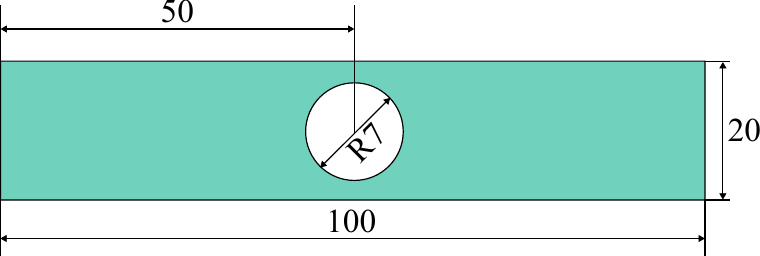}
       ~~~~ \includegraphics[width=0.45\textwidth]{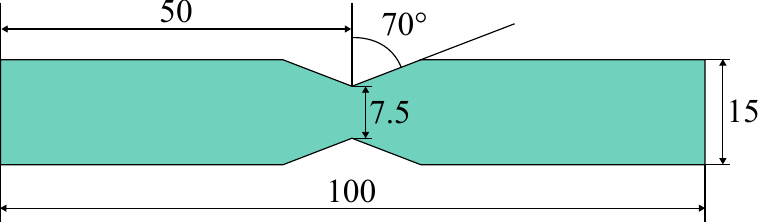}
    \caption{Geometry of the two-dimensional initial-boundary-value problems (IBVPs) defined in ABAQUS v.2022.}
    \label{fig:geometry_part}
\end{figure}

Two numerical examples are presented to examine the feasibility of the introduced approach: A two-dimensional plate with a hole and a two-dimensional plate with a double notch. These benchmark problems are expected to demonstrate the stress concentration effect for plasticity at the center of the plate, i.e. at the narrow areas, under monotonic tension loading conditions. The geometry of the samples is illustrated in Fig. \ref{fig:geometry_part}. 

Both plates are completely fixed on the left edge and subjected to a monotonic displacement of 0.01 on the right edge along the x-axis. Other degrees of freedom along the $y$- and $z$-axis are free. Both plates use a fine mesh with a global element size of 1.0, composed of triangular plane-stress CPS3 elements and quadrilateral elements CPS4. Automatic mesh refinement is implemented in the critical regions. The default element thickness is set to 1.0\,. For the current modeling, it is important to notice that the subjected displacement limit should not exceed the maximum value considered in the training data. This will help avoiding NN's inherent low accuracy in the extrapolation (larger errors for displacement limits out of the training range). 

In the case of the plate with a hole, the suggested initial increment size is chosen as 0.0014. The total simulation time is 92.88 seconds. In the case of the plate with a notch, the suggested initial increment size is chosen as 0.0012. The simulation time is 96.23 seconds.
To evaluate the validity of the proposed implementation, a comparison between the ML-based approach and a conventional plasticity model is presented. In this, the simulation results based on the ML model are compared with that of a simple {\it isotropic hardening plasticity model} provided in ABAQUS/Standard and characterized by a monotonic stress-strain curve. The Young's modulus and Poisson's ratio of the material are fitted from the generated database. The fitted Young's modulus was rounded to 1500, and the fitted Poisson's ratio was 0.33\,.

\subsection{Simulation resutls}
\begin{figure}[!h]
    \centering
    \begin{subfigure}[t]{0.75\textwidth}
        \includegraphics[width=\textwidth]{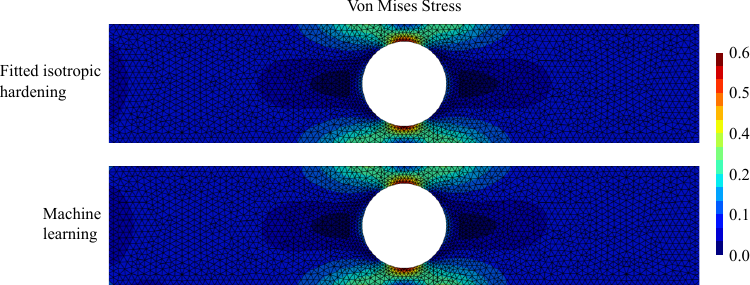}
        \caption{Geometry~(1): 2D plane-stress IBVP of a plate with a hole.}
        \label{fig:sub_plate_coarse}
    \end{subfigure}
    \par\bigskip
    \begin{subfigure}[t]{0.75\textwidth}
        \includegraphics[width=\textwidth]{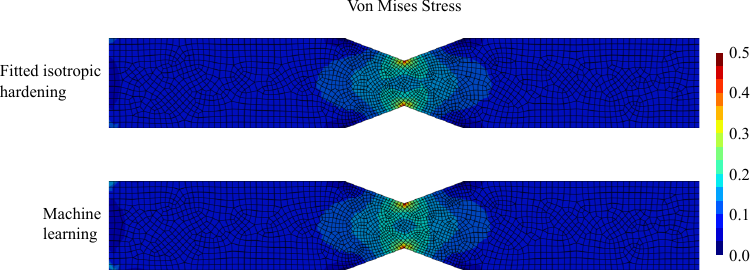}
        \caption{Geometry~(1):2D plane-stress IBVP of a plate with a double notch.}
        \label{fig:sub_plate_fine}
    \end{subfigure}
    \caption{Two initial-boundary-value problems (IBVPs) for comparison of the von Mieses stress components (in MPa) between the fitted isotropic hardening model and the ML-based material model in ABAQUS/Standard.}
    \label{fig:von_mises}
\end{figure}

Figure~\ref{fig:von_mises} illustrates the nodal results of the von~Mises stress after deformation for both cases of the data-based material model and the fitted isotropic hardening plasticity model of ABAQUS/Standard. Results indicate that the LSTM model effectively captures the stress concentration phenomenon around the hole and the notch. In the case of the plate with a hole, the maximum von~Mises stress was 0.6222 MPa, and the minimum was 0.0266 MPa. In the case of the plate with a notch, the maximum von~Mises stress was 0.4840 MPa, and the minimum was 0.0534 MPa. 

\subsection{Discussion of results and convergence}
\subsubsection{Impact of initial increment size}
To attain accurate results, it is important to ensure convergence during simulation. It is observed during the simulation process that the suggested initial value of increment size significantly influences the convergence performance for different displacement loads. 
\begin{figure}[htb]
\centering
\includegraphics[width=0.5\textwidth]{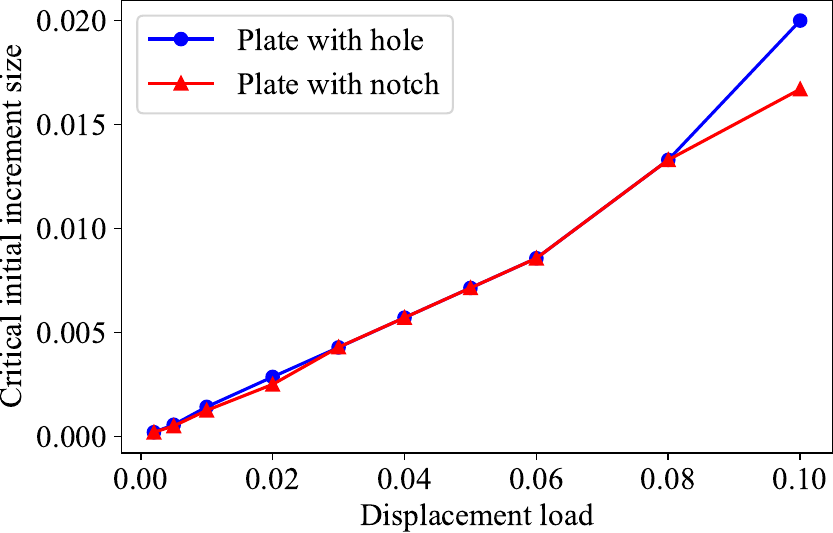}
\caption{Critical minimal incremental size related to displacement load within a single load step.}    \label{fig:critical_increment_size}
\end{figure}
Figure \ref{fig:critical_increment_size} shows the correlation between the critical minimal increment size in displacement and the overall displacement experienced within a single load step. These critical values indicate that, to achieve converged solutions, the suggested initial increment size for simulation should exceed the critical initial increment size. An initial increment size below the corresponding critical value leads to divergence and subsequent errors in the simulation result. 
\subsubsection{Impact of mesh fineness}
\begin{figure}[H]
    \centering
    \begin{subfigure}[b]{0.58\textwidth}
        \includegraphics[width=\textwidth]{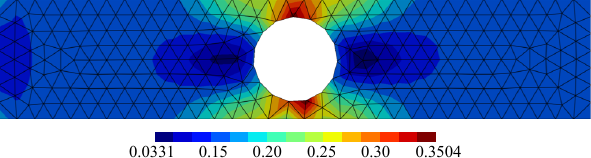}
        \caption{2D plat with hole, coarse mesh.}
    \end{subfigure}
    \par\bigskip
    \begin{subfigure}[b]{0.58\textwidth}
        \includegraphics[width=\textwidth]{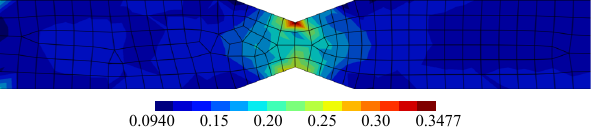}
        \caption{2D plate with a notch, coarse mesh.}
    \end{subfigure}
    \caption{Von Mises stress for models with coarse mesh. The initial displacement load increment size is identical to the fine mesh model.}
    \label{fig:coarse_mesh_mises}
\end{figure}

To investigate the impact of mesh fineness on the convergence behavior, simulations are conducted on the same part geometry employing a coarser mesh. The simulation settings, including the initial displacement increment size, remained unaltered. Figure \ref{fig:coarse_mesh_mises} shows the von~Mises stress distribution. The acquired results show that the mesh fineness notably influences result precision, while its impact on the convergence process itself is comparatively minor. 
%

\section{Summary and conclusions}
\label{sec:Conclusions}
This manuscript presents an approach to integrate machine learning-based material models for complex path-dependent constitutive relationships, such as crystal plasticity, into the finite element simulation framework of ABAQUS UMAT. The procedure particularly addresses solving nonlinear problems through the iterative Newton-Raphson method by substituting the material model with a neural network-based structure. The proposed methodology reconstructs offline-trained LSTM neural network calculation paths within ABAQUS, encapsulating path-dependent stress-strain relations at individual material points. Thus, it assists in improving the accuracy of describing nonlinear material responses. 
Two numerical 2D plane-stress IBVPs are conducted under uniaxial displacement loading conditions to prove the quality and applicability of the proposed approach.
It is found in this study that using an evenly distributed training dataset across the whole value range can improve the accuracy of NN's stress prediction. Moreover, using the direct graph approach instead of the informed graph can lower the training errors. It is also worth mentioning that NN prediction inherently contains bias and variance, which can lead to convergence challenges during analysis. In this connection, it is found that a displacement load below the critical value results in divergence. 

In conclusion, the integration of ML techniques offers an efficient and accurate method for modeling path-dependent material behaviors, such as crystal plasticity. This efficiency is primarily achieved because the trained ML model embedded within the UMAT behaves like an elasticity model during simulations. This eliminates the need for explicit determination of the material parameters, which is often required in traditional material models. In conventional models, material parameters such as elastic moduli, yield stress, and hardening rules must be determined either experimentally or through fitting processes, which can be time-consuming and difficult to generalize across different loading scenarios. However, with the ML approach, these parameters are implicitly learned from the data, simplifying the simulation setup.

The accuracy of this approach is rooted in the fact that the ML model is trained on a database generated from lower-scale simulations of crystal plasticity. By relying on lower-scale simulations that capture the detailed microscopic interactions and deformation mechanisms, such as slip behaviors in crystals, the model can make predictions that reflect the true material response with high precision. This ability to incorporate microstructural information allows the ML model to overcome the limitations of traditional constitutive models, which are often based on simplified assumptions and overlook the complexities of material behavior at lower scales.

Future research works will explore incorporating more complex material behavior in connection with ML-based techniques into the FE framework. This could involve, for example, modeling hydraulic fracturing with a phase-field approach \cite{Heider2021_review_PFHyd} or simulating additive manufacturing processes \cite{AliEtAl2023_AdditiveMan}.


\section*{Conflict of interest}
 The authors declare that there is no conflict of interest.

\section*{Data Availability}
Exemplary source codes are made available as open-access for interested readers at [\href{https://doi.org/10.25835/6n5uu50y}{https://doi.org/10.25835/6n5uu50y}]

\begin{acknowledgements}
The second author, Y. Heider, would like to gratefully thank the  German Research Foundation (DFG) for the support of the project  ``Multi-field continuum modeling of two-fluid-filled porous media fracture augmented by microscale-based machine-learning material laws'', grant number 458375627.
\end{acknowledgements}

%
%

\bibliographystyle{spbasic}      
\bibliography{main}

\begin{thebibliography}{42}
\providecommand{\natexlab}[1]{#1}
\providecommand{\url}[1]{{#1}}
\providecommand{\urlprefix}{URL }
\expandafter\ifx\csname urlstyle\endcsname\relax
  \providecommand{\doi}[1]{DOI~\discretionary{}{}{}#1}\else
  \providecommand{\doi}{DOI~\discretionary{}{}{}\begingroup
  \urlstyle{rm}\Url}\fi
\providecommand{\eprint}[2][]{\url{#2}}

\bibitem[{Abadi et~al.(2016)Abadi, Agarwal, Barham, Brevdo, Chen, Citro,
  Corrado, Davis, Dean, Devin et~al.}]{abadi2016tensorflow}
Abadi M, Agarwal A, Barham P, Brevdo E, Chen Z, Citro C, Corrado GS, Davis A,
  Dean J, Devin M, et~al. (2016) Tensorflow: Large-scale machine learning on
  heterogeneous distributed systems. arXiv preprint arXiv:160304467

\bibitem[{Aldakheel et~al.(2023{\natexlab{a}})Aldakheel, Elsayed, Tarek, and
  Wriggers}]{Aldakheel_2023_CNN_Hetero}
Aldakheel F, Elsayed E, Tarek Z, Wriggers P (2023{\natexlab{a}}) Efficient
  multiscale modeling of heterogeneous materials using deep neural networks.
  Comput Mech 72:155--171, \doi{10.1007/s00466-023-02324-9}

\bibitem[{Aldakheel et~al.(2023{\natexlab{b}})Aldakheel, Soyarslan, Palanisamy,
  and Elsayed}]{Aldakheel_2023_CNN_Magnetostatics}
Aldakheel F, Soyarslan C, Palanisamy H, Elsayed E (2023{\natexlab{b}}) Machine
  learning aided multiscale magnetostatics. arXiv preprint -:1--18,
  \doi{arXiv:2301.12782}

\bibitem[{Ali et~al.(2024)Ali, Heider, and Markert}]{AliEtAl2023_AdditiveMan}
Ali B, Heider Y, Markert B (2024) Predicting residual stresses in slm additive
  manufacturing using a phase-field thermomechanical modeling framework.
  Computational Materials Science 231:112576,
  \doi{https://doi.org/10.1016/j.commatsci.2023.112576}

\bibitem[{Borja(2013)}]{BORJA2013}
Borja RI (2013) Plasticity Modeling \& Computation. Springer Heidelberg New
  York Dordrecht London, \doi{10.1007/978-3-642-38547-6}

\bibitem[{Chaaban et~al.(2023)Chaaban, Heider, Sun, and
  Markert}]{ChaabanEtAl2023_ML_LBM}
Chaaban M, Heider Y, Sun W, Markert B (2023) A machine-learning supported
  multi-scale lbm-tpm model of unsaturated, anisotropic, and deformable porous
  materials. Int J Numer Anal Methods Geomech pp 1--22,
  \doi{https://doi.org/10.1002/nag.3668}

\bibitem[{Chollet et~al.(2015)}]{chollet2015keras}
Chollet F, et~al. (2015) Keras. \url{https://github.com/fchollet/keras}

\bibitem[{Dahmen and Reusken(2008)}]{DAHMEN2008}
Dahmen W, Reusken A (2008) Numerik für Ingenieure und Naturwissenschaftler.
  \doi{10.1007/978-3-540-76493-9}

\bibitem[{Dhillon and Verma(2020)}]{Dhillon_Verma_2020_reviewCNN}
Dhillon A, Verma G (2020) Convolutional neural network: a review of models,
  methodologies and applications to object detection. Prog Artif Intell
  9:85--112, \doi{https://doi.org/10.1007/s13748-019-00203-0}

\bibitem[{Eidel(2023)}]{EIDEL2023_CNN}
Eidel B (2023) Deep cnns as universal predictors of elasticity tensors in
  homogenization. Computer Methods in Applied Mechanics and Engineering
  403:115741

\bibitem[{Garcia-Teijeiro and
  Rodriguez-Herrera(2021)}]{TeijeiroEtAl2021_ML_FEM}
Garcia-Teijeiro X, Rodriguez-Herrera A (2021) {Combined Machine-Learning and
  Finite-Element Approach for Multiscale 3D Stress Modeling}. SPE Reservoir
  Evaluation \& Engineering 24(04):827--846, \doi{10.2118/205493-PA}

\bibitem[{Giovine(2022)}]{GIOVINE2022}
Giovine P (2022) Continua with partially constrained microstructure. Continuum
  Mechanics and Thermodynamics 34:273--295, \doi{10.1007/s00161-021-01057-5}

\bibitem[{Gorji et~al.(2020)Gorji, Mozaffar, Heidenreich, Cao, and
  Mohr}]{GORJI2020}
Gorji MB, Mozaffar M, Heidenreich JN, Cao J, Mohr D (2020) On the potential of
  recurrent neural networks for modeling path dependent plasticity. Journal of
  the Mechanics and Physics of Solids 143:103972,
  \doi{10.1016/j.jmps.2020.103972}

\bibitem[{Gu et~al.(2018)Gu, Wang, Kuen, Ma, Shahroudy, Shuai, Liu, Wang, Wang,
  Cai, and Chen}]{GU2018CNN}
Gu J, Wang Z, Kuen J, Ma L, Shahroudy A, Shuai B, Liu T, Wang X, Wang G, Cai J,
  Chen T (2018) Recent advances in convolutional neural networks. Pattern
  Recognition 77:354--377, \doi{https://doi.org/10.1016/j.patcog.2017.10.013}

\bibitem[{Guan et~al.(2023)Guan, Qu, Feng, Ma, and Zhou}]{Guan2023_FEM_ML}
Guan S, Qu T, Feng YT, Ma G, Zhou W (2023) A machine learning-based multi-scale
  computational framework for granular materials. Acta Geotech 18:1699–1720,
  \doi{https://doi.org/10.1007/s11440-022-01709-z}

\bibitem[{Hashash et~al.(2004)Hashash, Jung, and Ghaboussi}]{HASHASH2004}
Hashash YMA, Jung S, Ghaboussi J (2004) Numerical implementation of a neural
  network based material model in finite element analysis. International
  Journal for Numerical Methods in Engineering 59:989--1005,
  \doi{10.1002/nme.905}

\bibitem[{Heider(2021{\natexlab{a}})}]{Heider2021_Habil}
Heider Y (2021{\natexlab{a}}) Multi-field and multi-scale computational
  fracture mechanics and machine-learning material modeling. Habilitation,
  Report No. IAM-13 of Institute of General Mechanics, RWTH Aachen University

\bibitem[{Heider(2021{\natexlab{b}})}]{Heider2021_review_PFHyd}
Heider Y (2021{\natexlab{b}}) A review on phase-field modeling of hydraulic
  fracturing. Engineering Fracture Mechanics 253:107881

\bibitem[{Heider and Sun(2023)}]{HeiderSun2023_PAMM_PlasObj}
Heider Y, Sun W (2023) Objectivity and accuracy enhancement within ann-based
  multiscale material modeling. PAMM 22(1):e202200203,
  \doi{https://doi.org/10.1002/pamm.202200203}

\bibitem[{Heider et~al.(2020)Heider, Wang, and Sun}]{HEIDER2020}
Heider Y, Wang K, Sun W (2020) {{SO}}(3)-invariance of informed-graph-based
  deep neural network for anisotropic elastoplastic materials. Computer Methods
  in Applied Mechanics and Engineering 363:112875,
  \doi{10.1016/j.cma.2020.112875}

\bibitem[{Heider et~al.(2021)Heider, Suh, and Sun}]{HeiderHSSuh2020_ML_offline}
Heider Y, Suh HS, Sun W (2021) An offline multi-scale unsaturated poromechanics
  model enabled by self-designed/self-improved neural networks. Int J Numer
  Anal Methods Geomech 45(9):1212--1237, \doi{https://doi.org/10.1002/nag.3196}

\bibitem[{Heider et~al.(2024)Heider, Aldakheel, and
  Ehlers}]{HeiderAldakheelEhlers_2024CNN}
Heider Y, Aldakheel F, Ehlers W (2024) Cnn-powered micro- to macro-scale flow
  modeling in deformable porous media. Meccanica -:under review

\bibitem[{Hochreiter and Schmidhuber(1997)}]{HOCHREITER1997}
Hochreiter S, Schmidhuber J (1997) {{LONG SHORT-TERM MEMORY}}. Neural
  Computation 9:1735--1780, \doi{10.1162/neco.1997.9.8.1735}

\bibitem[{Huang et~al.(2020)Huang, Fuhg, Wei{\ss}enfels, and
  Wriggers}]{HUANG2020a}
Huang D, Fuhg JN, Wei{\ss}enfels C, Wriggers P (2020) A machine learning based
  plasticity model using proper orthogonal decomposition. Computer Methods in
  Applied Mechanics and Engineering 365:113008, \doi{10.1016/j.cma.2020.113008}

\bibitem[{Jang et~al.(2021)Jang, Fazily, and Yoon}]{JANG2021102919_UMAT}
Jang DP, Fazily P, Yoon JW (2021) Machine learning-based constitutive model for
  j2- plasticity. International Journal of Plasticity 138:102919,
  \doi{https://doi.org/10.1016/j.ijplas.2020.102919}

\bibitem[{Koeppe et~al.(2020)Koeppe, Bamer, and
  Markert}]{KoeppeBamerMarkert2020_An_Intelligent}
Koeppe A, Bamer F, Markert B (2020) An intelligent nonlinear meta element for
  elastoplastic continua: deep learning using a new time-distributed residual
  u-net architecture. Computer Methods in Applied Mechanics and Engineering
  366:113088, \doi{https://doi.org/10.1016/j.cma.2020.113088}

\bibitem[{Lazar(2009)}]{LAZAR2009}
Lazar A (2009) {{SORN}}: A {{Self-organizing Recurrent Neural Network}}.
  Frontiers in Computational Neuroscience 3, \doi{10.3389/neuro.10.023.2009}

\bibitem[{Lee et~al.(2021)Lee, Wu, Lin, and Ko}]{Lee2021_ML_FEM}
Lee YT, Wu TH, Lin ML, Ko CC (2021) Machine (Deep) Learning and Finite Element
  Modeling, Springer International Publishing, Cham, pp 183--188.
  \doi{10.1007/978-3-030-71881-7_14}

\bibitem[{Lefik and Schrefler(2003)}]{LEFIK2003}
Lefik M, Schrefler B (2003) Artificial neural network as an incremental
  non-linear constitutive model for a finite element code. Computer Methods in
  Applied Mechanics and Engineering 192:3265--3283,
  \doi{10.1016/S0045-7825(03)00350-5}

\bibitem[{Li et~al.(2019)Li, Roth, and Mohr}]{LI2019}
Li X, Roth CC, Mohr D (2019) Machine-learning based temperature- and
  rate-dependent plasticity model: {{Application}} to analysis of fracture
  experiments on {{DP}} steel. International Journal of Plasticity
  118:320--344, \doi{10.1016/j.ijplas.2019.02.012}

\bibitem[{Lourenço et~al.(2022)Lourenço, Andrade-Campos, and
  Georgieva}]{LourencoEtAl2022_VMAT_met12030427}
Lourenço R, Andrade-Campos A, Georgieva P (2022) The use of machine-learning
  techniques in material constitutive modelling for metal forming processes.
  Metals 12(3), \doi{10.3390/met12030427},
  \urlprefix\url{https://www.mdpi.com/2075-4701/12/3/427}

\bibitem[{Maia et~al.(2023)Maia, Rocha, Kerfriden, and Van Der~Meer}]{MAIA2023}
Maia M, Rocha I, Kerfriden P, Van Der~Meer F (2023) Physically recurrent neural
  networks for path-dependent heterogeneous materials: {{Embedding}}
  constitutive models in a data-driven surrogate. Computer Methods in Applied
  Mechanics and Engineering 407:115934, \doi{10.1016/j.cma.2023.115934}

\bibitem[{Miyazawa et~al.(2019)Miyazawa, Briffod, Shiraiwa, and
  Enoki}]{MIYAZAWA2019}
Miyazawa Y, Briffod F, Shiraiwa T, Enoki M (2019) Prediction of {{Cyclic
  Stress}}\textendash{{Strain Property}} of {{Steels}} by {{Crystal Plasticity
  Simulations}} and {{Machine Learning}}. Materials 12:3668,
  \doi{10.3390/ma12223668}

\bibitem[{Oishi and Yagawa(2017)}]{OISHI2017}
Oishi A, Yagawa G (2017) Computational mechanics enhanced by deep learning.
  Computer Methods in Applied Mechanics and Engineering 327:327--351,
  \doi{10.1016/j.cma.2017.08.040}

\bibitem[{Pantidis and Mobasher(2023)}]{PANTIDIS2023}
Pantidis P, Mobasher ME (2023) Integrated {{Finite Element Neural Network}}
  ({{I-FENN}}) for non-local continuum damage mechanics. Computer Methods in
  Applied Mechanics and Engineering 404:115766, \doi{10.1016/j.cma.2022.115766}

\bibitem[{Patra et~al.(2023)Patra, Dhar, and Acharyya}]{PatraEtAl2023_FE_ANN}
Patra S, Dhar S, Acharyya S (2023) Finite element implementation of ann-based
  constitutive models for dsa in ss304. J Inst Eng India Ser D pp~--,
  \doi{https://doi.org/10.1007/s40033-023-00475-w}

\bibitem[{Suh et~al.(2023)Suh, Kweon, Lester, Kramer, and
  Sun}]{Suh2023_PyTorch_ABAQUS_Umat}
Suh HS, Kweon C, Lester B, Kramer S, Sun W (2023) A publicly available
  pytorch-abaqus umat deep-learning framework for level-set plasticity.
  Mechanics of Materials 184:104682,
  \doi{https://doi.org/10.1016/j.mechmat.2023.104682},
  \urlprefix\url{https://www.sciencedirect.com/science/article/pii/S016766362300128X}

\bibitem[{Tandale and Stoffel(2023)}]{Tandale_2023_CNN_RNN}
Tandale S, Stoffel M (2023) Recurrent and convolutional neural networks in
  structural dynamics: a modified attention steered encoder–decoder
  architecture versus lstm versus gru versus tcn topologies to predict the
  response of shock wave-loaded plates. Comput Mech
  \doi{10.1007/s00466-023-02317-8}

\bibitem[{Tao et~al.(2021)Tao, Liu, Du, and Yu}]{TAO2021}
Tao F, Liu X, Du H, Yu W (2021) Learning composite constitutive laws via
  coupling {{Abaqus}} and deep neural network. Composite Structures 272:114137,
  \doi{10.1016/j.compstruct.2021.114137}

\bibitem[{Weber et~al.(2023)Weber, Wagner, and
  Freitag}]{WeberWagnerFreitag2023}
Weber P, Wagner W, Freitag S (2023) Physically enhanced training for modeling
  rate-independent plasticity with feedforward neural networks. Comput Mech
  72:827--857, \doi{https://doi.org/10.1007/s00466-023-02316-9}

\bibitem[{Xu et~al.(2021)Xu, Huang, and Darve}]{XU2021}
Xu K, Huang DZ, Darve E (2021) Learning {{Constitutive Relations}} using
  {{Symmetric Positive Definite Neural Networks}}. Journal of Computational
  Physics 428:110072, \doi{10.1016/j.jcp.2020.110072}

\bibitem[{Yuan et~al.(2018)Yuan, Paradiso, Meredig, and Niezgoda}]{YUAN2018}
Yuan M, Paradiso S, Meredig B, Niezgoda SR (2018) Machine
  {{Learning}}\textendash{{Based Reduce Order Crystal Plasticity Modeling}} for
  {{ICME Applications}}. Integrating Materials and Manufacturing Innovation
  7:214--230, \doi{10.1007/s40192-018-0123-x}

\end{thebibliography}
%
%

\end{document}